\documentclass{svproc}
\usepackage{url}

\usepackage{graphicx}
\usepackage{siunitx}
\usepackage{float} 
\usepackage{amsmath}
\usepackage{amsfonts}
\usepackage{wasysym} 

\begin{document}

\mainmatter              
%
%
\title{Exploring Coevolutionary Dynamics\\ 
of Competitive Arms-Races
Between\\ Infinitely Diverse Heterogenous\\Adaptive Automated Trader-Agents
}
\titlerunning{Coevolutionary Dynamics of Trader-Agents}  
%
\author{Nik Alexandrov \and Dave Cliff  \and Charlie Figuero\footnote{Authors in alphabetic order. Direct all correspondence to Dave Cliff. 
\\ Copyright \copyright 2021 N. Alexandrov, D. Cliff, \& C. Figuero.\\
{\em Accepted for publication at the 16th Annual Social Simulation Conference, Krakow, Poland, 20--24 September 2021.}
}}
\authorrunning{Alexandrov, Cliff, \& Figuero} 
%
\tocauthor{Nik Alexandrov, Dave Cliff, Charlie Figuero}
\institute{Department of Computer Science, University of Bristol, Bristol BS8 1UB, U.K.\\
\email{na17807@bristol.ac.uk}, \email{csdtc@bristol.ac.uk}, \email{cf17559@bristol.ac.uk}
}

\maketitle              


\begin{abstract}
%
%
We report on a series of experiments in which we study the coevolutionary ``arms-race'' dynamics among groups of agents that engage in adaptive automated trading in an accurate model of contemporary financial markets. 
At any one time, every trader in the market is trying to make as much profit as possible given the current distribution of different {\em other}\/ trading strategies that it finds itself pitched against in the market; but the distribution of trading strategies and their observable behaviors is constantly changing, and changes in any one trader are driven to some extent by the changes in all the others. Prior studies of coevolutionary dynamics in markets have concentrated on systems where traders can choose one of a small number of fixed pure strategies, and can change their choice occasionally, thereby giving a market with a discrete phase-space, made up of a finite set of possible system states. Here we present first results from two independent sets of experiments, where we use minimal-intelligence trading-agents but in which the space of possible strategies is continuous and hence infinite. Our work reveals that by taking only a small step in the direction of increased realism we move immediately into high-dimensional phase-spaces, which then present difficulties in visualising and understanding the coevolutionary dynamics unfolding within the system. We conclude that further research is required to establish better analytic tools for monitoring activity and progress in co-adapting markets. We have released relevant Python code as open-source on GitHub, to enable others to continue this work.
\end{abstract}

\section{Introduction}
In the past 20 years most of the world's major financial markets have seen a sharp rise in the level of automated trading on those markets, with many human traders being replaced by adaptive algorithmic ``robot traders'' at the point of execution. Although this has been a significant shift, affecting both patterns of employment and the dynamics of the markets concerned, it can plausibly be argued that at a macro-level little has changed: these major markets are still populated by traders working on behalf of major financial institutions such as investment banks or fund-management companies; the difference is just that now those institutions are represented in the markets not by teams of human traders but by teams of robots. To be more precise, it is more often the case that within any one institution entire teams of human traders have been replaced by a single monolithic automated trading system that does the work previously performed by tens of hard-working human traders. 

The success or failure of any one automated trading system is determined primarily by how much profit it can generate, but underlying that simple observation is a circularity. In any realistic market scenario, the profitability of a given robot trader $R_1$ will be determined at least in part by the extent to which its actions in the market are well-tuned to the likely reactions of other traders in that market. Thus, in contemporary markets, $R_1$ is likely to be designed to adapt its trading behavior to the current market circumstances, and yet those circumstances are significantly determined by the behavior of other traders in the market, most of which are robots  $R_2$, $R_3$, $R_4$ and so on, each of which are themselves adapting to the circumstances they experience in the market, which are to some extent influenced by the actions and reactions of $R_1$. 

In the natural world, in the Darwinian survival-of-the-fittest interactions among evolving species of organisms, exactly this kind of circular interaction and dependency is commonplace.
Just as the profit-driven adaptive trading behavior of robot $R_1$ can be affected by the profit-driven adaptive trading behavior of $R_2$, and {\em vice versa}, so the reproductive fitness of a predator animal (a cheetah, say) is determined to some extent by how well adapted it is to catching its prey, and the reproductive fitness of individuals in its prey species (antelopes, say) is starkly dependent on the extent to which they are adapted to evade being caught by their predators. If a mutation in the predator species gives rise to individuals that can run faster for longer when chasing prey, perhaps that will subsequently be countered by the prey species evolving to turn more sharply or to jump higher or farther than the predator can deal with. Similarly if the prey species happens to evolve sharper eyesight so they can better see the predator coming, perhaps the predator species will then evolve to exhibit stealthier ways of tracking their food. This circular arms-race dynamic, where evolutionary adaptations in one species $S_1$ are driven by the current distribution of genes in one or more other species $S_2$, $S_3$, and so on, and where in turn evolutionary adaptations in one or more of those other species $S_2$, $S_3$ etc are driven by the current distribution of genes in species $S_1$, 
is known technically as {\em coevolution}. Theoretical biologists have studied coevolution for many years, and have developed various game-theoretical analyses that give insights on the dynamics of the arms-races between competitively coevolving species: see e.g.\ 
\cite{maynardsmith_1982,kauffman_1993,thompson_1994,hebbron_etal_2008}).

In this paper we report on empirical simulation studies for which the starting-point draws direct inspiration from those theoretical biology studies of coevolutionary dynamics. Our motivation here is to try to better understand, to gain insights into, the practical extent to which the various adaptive trading systems in a market are affecting each other, and specifically to investigate whether the population of adaptive traders is ever likely to converge on a situation where all of the traders are well-adapted to each others' behavior yet each trader is not as profitable as it could otherwise be. That is: could the competitive interactions and adaptations of traders in the market collectively converge on a stable set of trading behaviors that are sub-optimal? And, if so, can we recognize when that has happened, or when it is about to happen? 
Similarly, might we be able to identify when the coevolutionary dynamics are about to lead to a flash-crash?
We have commenced a sequence of empirical studies, starting with minimal but realistic simulations that are principled approximations of present-day highly automated financial markets. 

Specifically, our ultimate aim has been to create agent-based models (ABMs) involving $N_A$ agents where each agent represents one financial {\em legal entity} (i.e. either an individual independent trader or an institution such as a bank or fund-management company) operating a single profit-driven automated trading system that trades in competition with the $N_A-1$ other agents, in an electronic market operating a continuous double auction (CDA) with a limit order book (LOB: see e.g.\ 
\cite{gould_etal_2013_LOBs,nolte_etal_2014,abergel_etal_2016})  
-- which is the present-day situation in many of the world's major financial markets. Each entity can in principle be adapting its strategy/behavior in real-time (e.g.\ using a machine learning mechanism) but is not required to do so. That is, an entity's trading strategy can be non-adaptive if that is the more profitable option. Furthermore at any time the entity can elect to totally change the strategy that it is operating, modelling the case where a financial institution switches a trading algorithm that has previously been in development and testing (commonly referred to as a {\em dev algo}) into full use, the dev algo replacing the previously-running {\em production algorithm} (commonly known as the {\em prod algo}). Thus each trading entity in our model internally maintains a minimum of two strategies, each of which could potentially be adaptive: a prod algo and a dev algo. When the agent's dev algo replaces the prod, a new dev algo is created and is subsequently tested and refined until there is sufficient evidence that it is an improvement on the agent's current prod algo, at which point the dev again replaces prod and then another new dev is created. The trade-off between {\em exploiting}\/ the prod algo and {\em exploring}\/ the dev algo has manifest links to studies of multi-armed bandit problems (see e.g.\ 
\cite{myleswhite_2012,lattimore_szepesvari_2020,slivkins_2021}).
We report here on the construction of two simulation models of this kind of system and on results from hundreds of thousands of simulated market sessions.

Simulation modelling of financial markets very often involves populating a market mechanism with some number of {\em trader-agents}: autonomous entities that have ``agency'' in the sense that they are empowered to buy and/or sell items within the particular market mechanism that is being simulated: this approach, known as {\em agent-based computational economics} 
(ACE: see e.g. \cite{hommes_lebaron_2018}), 
has a history stretching back for more than 30 years, and much of the work in ACE studies of trading behaviours in models of financial markets owes a clear intellectual debt to work in experimental economics as pioneered by Vernon Smith (see e.g.\ 
\cite{smith_1962,smith_1991,kagel_roth_1997,smith_2000,plott_smith_2008}). 

Over the multi-decade history of ACE, a small number of specific trader-agent algorithms, i.e. precise mathematical and algorithmic specifications of particular trading strategies, have been frequently used for modelling various aspects of financial markets, and the convention that has emerged is to refer to each such strategy via a short sequence of letters, reminiscent of a stock-market ticker-symbol. Notable trading strategies in this literature include (in chronological sequence): SNPR \cite{rust_etal_1992}, ZIC \cite{gode_sunder_1993}, ZIP \cite{cliff_1997_zip}, GD \cite{gjerstad_dickhaut_1998}, RE \cite{erev_roth_1998}, MGD \cite{tesauro_das_2001}, GDX \cite{tesauro_bredin_2002}, HBL \cite{gjerstad_2003}, and AA \cite{vytelingum_etal_2008}; several of which are explained in more detail later in this paper. Of these, ZIC (invented by the economists Gode \& Sunder \cite{gode_sunder_1993}) is notable for being both highly stochastic and extremely simple, and yet it gives surprisingly human-like market dynamics; GD  and ZIP were the first two strategies to be demonstrated as superior to human traders, a fact established in a landmark paper by IBM researchers \cite{das_etal_2001},
(see also: \cite{deluca_cliff_2011_icaart,deluca_cliff_2011_ijcai,deluca_etal_2011_foresight}), 
which is now commonly pointed to as initiating the rise of algorithmic trading in real financial markets; and until very recently AA was widely considered to be the best-performing strategy in the public domain. 
ZIC was the first instance of a {\em zero intelligence}\/ trading strategy, which have proven to be surprisingly useful in ACE research: see, e.g., 
\cite{farmer_etal_2005,ladley_2012}.  
With the exception of SNPR and ZIC, all later strategies in this sequence are adaptive, using some kind of machine learning (ML) or artificial intelligence (AI) method to modify their responses over time, better-fitting their trading behavior to the specific market circumstances that they find themselves operating in, and details of these algorithms were often published in major AI/ML conferences and journals.

The supposed dominance of AA has recently been questioned in a series of publications \cite{vach_2015,cliff_2019,snashall_cliff_2019,rollins_cliff_2020,cliff_rollins_2020} 
which demonstrated AA to have been less robust than was previously thought. Most notably, 
\cite{rollins_cliff_2020,cliff_rollins_2020} 
report on trials where AA is tested against two minimally simple algorithms that each involve no AI or ML at all: these two strategies are known as GVWY and SHVR \cite{cliff_2012_bse,cliff_2018_bse}, and each share the pared-back minimalism of Gode \& Sunder's ZIC mechanism. In the studies that have been published thus far, depending on the circumstances, it seems (surprisingly) that GVWY and SHVR can each outperform not only AA but also many of the other AI/ML-based trader-agent strategies in the set listed above. Given this surprising recent result, there is an appetite for further ACE-style market-simulation studies involving GVWY and SHVR. One compelling issue to explore is the coevolutionary dynamics of markets populated by traders that can choose to play one of the three strategies from GVWY, SHVR, and ZIC, in a manner similar to that studied by \cite{walsh_etal_2002} who employed {\em replicator dynamics}\/ modelling techniques borrowed from theoretical evolutionary biology to explore the coevolutionary dynamics of markets populated by traders that could choose between SNPR, ZIP, and GD; each trader playing their chosen strategy for as long as it seems (to that trader) to be the most profitable strategy, and occasionally switching to (or ``replicating'') use one of the other two strategies in the set if the current strategy appears (to that trader) to be weak. 
This replicator dynamics approach was also used in \cite{vytelingum_etal_2008} to argue that AA was dominant over prior leading strategies, and in \cite{vach_2015} to demonstrate that AA could in fact be dominated by other strategies. 

Replicator dynamics studies are typically limited to visualising and analysing the coevolutionary dynamics of simple, restricted systems where the restrictions are introduced to constrain the systems in such a way that they can be easily visualised and analysed. For instance, replicator dynamics studies often involve studying a population of agents that can switch between two, three, or at most four distinct pure strategies, and this decision often seems driven by the fact that visualisation of the dynamics, characterising the entire system dynamics, is often best done by reference to the system's {\em phase space}, i.e.\ to plot some factor of interest for every possible state of the system. Let $\cal S$ be the set of distinct pure strategies that the agents in our system can choose between, let $N_S = |{\cal S}|$ and  $s_i : i \in \{1,\ldots, N_s\}$ refer to the $i^{\text th}$ of those strategies. Also let $N_A$ be the number of agents in the system, each of which makes a choice of some $s_i \in {\cal S}$. Such a system can be characterised in full, all possible points in its finite phase space enumerated and plotted, by considering each possible combination of allowable strategy choices or assignments made by the population of agents: if all the $N_A$ agents have the same choice, and each can choose any of the $N_S$ strategies, then the number of possible system states, the number of points in its phase space, is $N_S^{N_A}$, a number that may grow large but will forever be finite.
  
When $N_S=2$, the system phase space can be characterised as points on a line, spanning from all agents playing $s_1$, through to a 50:50 mix of $s_1$:$s_2$, to all agents playing $s_2$.  When $N_S=3$, the phase space can be characterised and visualised as points on the {2D unit simplex}, an equilateral triangle where a point within or on the perimeter of the triangle represents a particular ratio of $s_1$:$s_2$:$s_3$, plotted in a barycentric coordinate frame. Technically, the one-dimensional (1D) line used for the phase-space of a $N_S=2$ system is a 2D unit simplex; the 3D unit simplex is a 2D triangle; and then the 4D unit simplex is a 3D object, a tetrahedron, the volume bounded by four planar faces each being an equilateral triangle. Higher-dimensional simplices are mathematically well-formed objects, but they are devils to visualise: try plotting the 40D unit simplex. Although the original authors do not explicitly state their reasons, it seems reasonable to conclude that each of \cite{walsh_etal_2002,vytelingum_etal_2008} and \cite{vach_2015} chose to study replicator dynamics systems in which $N_S=3$ and not any higher number because of the rapidly escalating difficulty of visualising the phase space for any higher value. Yet real-world markets do not involve all entities each selecting from a choice of two or three pure trading strategies, so there is then a major concern over the extent to which these studies adequately capture the much richer degree of heterogeneity in real-world markets: this brings to mind the old adage about the late-night drunkard looking for his lost house-keys under a streetlamp not because that is where he mislaid them, but because the light is better there. 

So, although one way of studying coevolutionary dynamics in markets where the traders can choose to either deploy GVWY, SHVR, or ZIC is to give each trader a discrete choice of one from that set of three strategies, so at any one time any individual trader is either operating according to GVWY or SHVR or ZIC, it is appealing to instead design experiments where the traders can continuously vary their trading strategy, exploring a potentially {\em infinite} range of differing trading strategies, where the space of possible strategies includes GVWY, SHVR, and ZIC. This is made possible by the recent introduction of a new minimal-intelligence trading strategy called PRZI \cite{cliff_2021_przi}. PRZI's trading behavior is determined by a strategy parameter $s \in [-1,+1] \in {\mathbb R} $. When $s=0$, the trader behaves identically to ZIC, and when $s = \pm 1$ it behaves the same as GVWY or SHVR. And, crucially, when a PRZI trader's $s$-value is some other value, either part-way between $-1$ and $0$ or part-way between $0$ and $+1$, its trading behavior is a kind of hybrid, part-way between that of ZIC and SHVR, or part-way between ZIC and GVWY.  Because the PRZI strategy-parameter $s$ is a real number, and its effect on the trading behavior is smooth and continuous, in principle any one PRZI trader can make microscopically small adjustments and hence the space of possible strategies available to a single PRZI trader is infinite, and the phase-space of a market of $N_A$ agents is a bounded volume within ${\mathbb R}^{N_A}$.  

In Section~\ref{sec:PRZI} we discuss our experiences in working with populations of coevolving PRZI traders, where we immediately come up against the limits of applicable visualisation techniques for this type of dynamical system. While markets of PRZI traders allow for continuous and infinite heterogeneity in the population of agents, the bounded nature of the PRZI strategy-space is a limitation that reduces the realism of the model. To address this, we have commenced work on an unboundedly infinite system, where each coevolving trader's strategy can in principle grow to be arbitrarily complex and sophisticated (that is, in principle they can be anything that is expressible as a program in a Turing-complete list-based functional programming language), which we discuss in Section~\ref{sec:STGP}. For all our simulation studies reported here, we use the BSE simulator of a CDA market with a LOB (see  \cite{cliff_2012_bse,cliff_2018_bse}), a mature open-source platform for ACE studies of electronic markets with automated trading. 

\section{Coevolution in a Bounded Infinite Space: PRZI}
\label{sec:PRZI}

Full details of our initial work with coevolving populations of PRZI traders are given in \cite{alexandrov_2021}, which this section is only a very brief summary of.

As a first illustration, we set up a minimal coevolutionary system, one in which only two of the traders could change their strategy by altering their PRZI $s$-value. Let's refer to these two traders as $T_1$ and $T_2$: the two are independent, so $T_1$ can set its strategy value $s_1$ regardless of the value $s_2$ chosen by $T_2$, and {\em vice versa}. We set to $T_1$ to be a buyer, and we set $T_2$ to be a seller and hence, because any seller in the market needs to find a buyer as a counter-party and {\em vice versa}, the profitability of $T_1$'s choice of $s_1$ will be partially dependent on $T_2$'s choice of $s_2$, and {\em vice versa}. We take the natural step of treating profitability as `fitness' in the evolutionary sense, and hence this system is as simple as we can get while still being coevolutionary. 

For the adaptation process, each adaptive trader operates a simple Adaptive Climber (AC) algorithm defined in \cite{alexandrov_2021}, which echoes the dev/prod development cycle discussed in the previous section: the trader maintains two separate strategies, to different PRZI $s$-values, referred to as $P$ and $D$. $P$ is initially set to some value, and $D$ is set to a `mutated' version of $P$, by adding a small random value (e.g. a sample from a uniform distribution over the range $[-0.05,+0.05]$). The AC method executes some number $N_T$ trades using strategy $P$ and then executes $N_T$ trades using strategy $D$. After that, if the profitability of $P$ is greater than that of $D$ then the trader generates a new $D$; but if the profitability of $D$ exceeds that of $P$ then $D$ is used to replace $P$, and then a new $D$ is generated as a mutant value of the new $P$. That is, AC is a minimally simple two-point stochastic hill-climber algorithm.

\begin{figure}
\begin{center}
\includegraphics[width=0.60\linewidth]{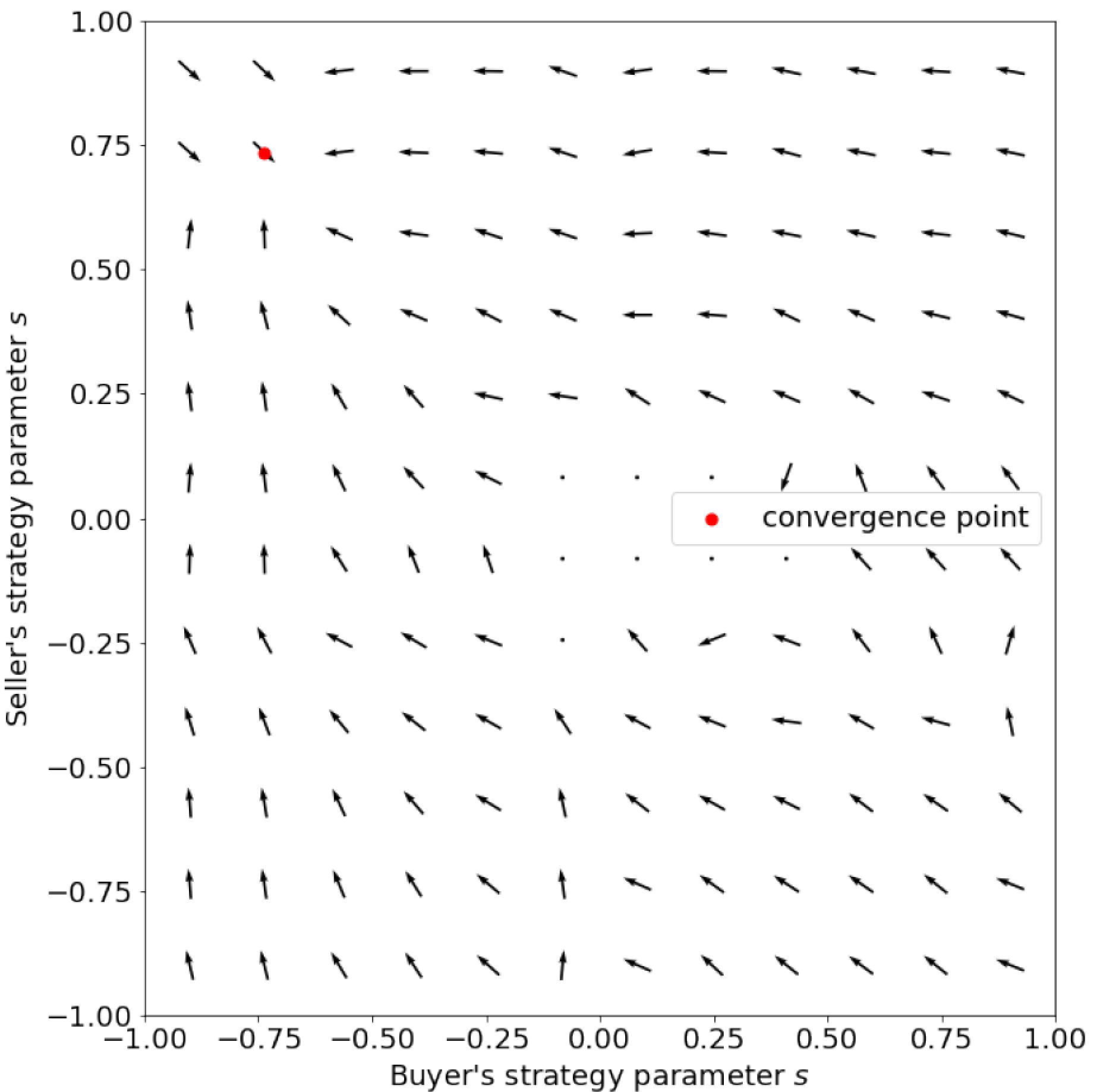}
\end{center}
\caption{Quiver plot over the phase-space for a minimal coevolutionary market populated wholly by PRZI traders in which only two of the traders are each independently adapting their $s_i$ strategy-values, while all other traders in the market hold their $s_i$ values constant. See text for further explanation and discussion. 
}
\label{fig:2dvectorfield}
\end{figure}

Figure~\ref{fig:2dvectorfield} shows a quiver plot of the phase space for an instance of this system, in which initial values of $s_i$ are set at random from a uniform distribution for all traders. One of the two adaptive traders is designated as a buyer with strategy $s_b(t)$, and the other as a seller with strategy $s_s(t)$. Both the buyer and the seller can adjust their strategy value over time, using the AC method just described. The horizontal axis is the buyer's $s_b$ value and the vertical is the sellers's $s_s$, and these two continuous values define the system's phase-space, i.e. $[-1,+1]^2 \in {\mathbb R}^2$. Uniform-length vectors have been plotted at regular intervals giving a discrete grid that indicates the system's direction of travel in phase-space. The phase-space has a single point-attractor, the point of convergence marked by a red dot at $s_b\approx-0.75, s_s\approx0.75$, and an obvious plateau area close to the origin: within the plateau area, the system will exhibit random drift, and will eventually step outside the plateau; once outside the plateau, the system evolves toward the attractor. 

The 2D quiver plot in Figure~\ref{fig:2dvectorfield} 
is made possible because we constrained our system to only have two adaptive traders. 
As soon as we relax that constraint and have all $N_A$ agents in our system adapting and coevolving against all the others, we need to make an $N_A$-dimensional plot. Given that we routinely use $N_A$ values of 50 or more, and that 50-dimensional quiverplots are not easy to plot or understand, this mode of visualization runs out of steam as soon as $N_A$ gets to plausibly interesting numbers.

An $N_A$-dimensional coevolutionary market system is an instance of an $N_A$-dimensional dynamical system, and a popular method of characterising the dynamics of high-dimensional dynamical systems is the {\em recurrence plot} (RP: see e.g.\
\cite{eckmann_etal_1987,marwan_2011}). 
This purely graphical technique can be extended by various quantitative methods known collectively as {\em recurrence quantification analysis} (RQA: see \cite{webber_marwan_2015}). As is discussed at length in \cite{alexandrov_2021}, we have explored the use of RPs and RQA for visualising and analysing our coevolutionary PRZI markets. 

\begin{figure}
\begin{center}
\includegraphics[width=0.40\linewidth]{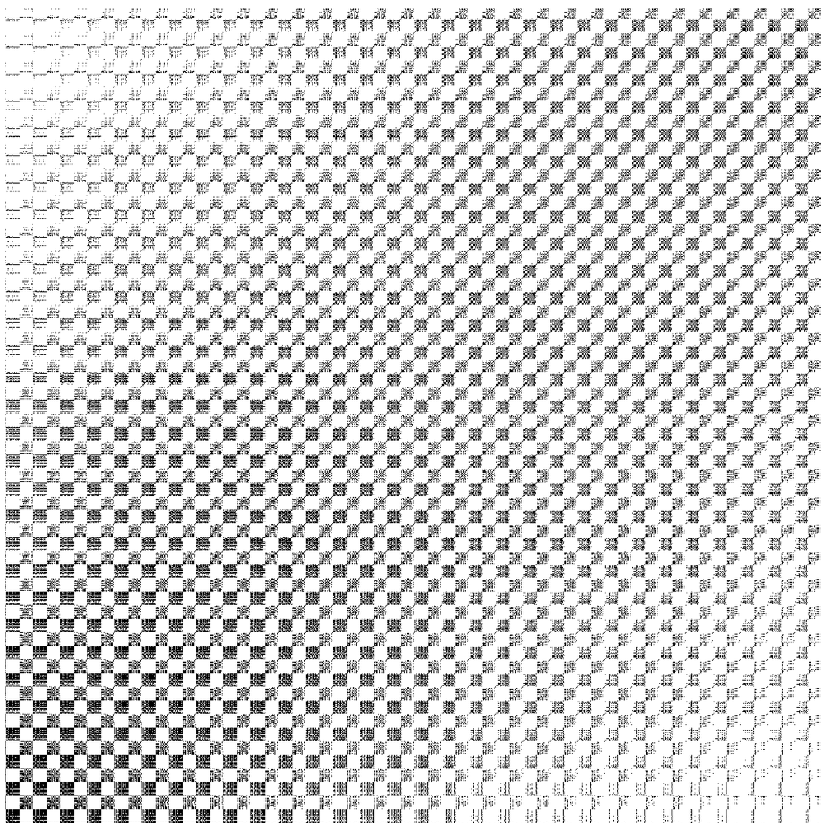}
\end{center}
\caption{Recurrence plot for a market of coevolving PRZI traders in which the number of adaptive traders $N_A >> 2$. See text for further explanation. 
}
\label{fig:recplot}
\end{figure}

In brief, for our purposes a RP visualization of an $N$-dimensional real-valued dynamical system is a rectangular grid of square binary pixels, i.e. pixels that are in one of two states: often either black or white. Let $\overrightarrow{s}(t) \in {\mathbb R}^N$ be the state of the system at time $t$. A pixel is shaded black to represent that  $\overrightarrow{s}(t)$ has {\em recurred} at time $t$, i.e.\ has previously been seen at some earlier time $t-\Delta_t$, and is shaded white otherwise. Recurrence can be defined in various ways, but the simplest is to take the $N$-dimensional Euclidian distance $d= |\overrightarrow{s}(t) - \overrightarrow{s}(t-\Delta_t)|$ and to declare recurrence to have occurred if $d$ is less than some threshold value. The co-ordinates for each pixel, 
each cell, 
in a RP are set by its values of $t$ and $t-\Delta_t$. Figure~\ref{fig:recplot} shows a RP for one instance of our coevolving market of PRZI traders: there is nontrivial structure in the plot, which is subjected to further detailed RQA analysis in \cite{alexandrov_2021}. 

In our work with coevolving PRZI traders, merely by allowing each zero-intelligence trader to have adaptive control of its single real-valued strategy parameter, for a market populated by $N_A$ such traders, we have an $N_A$-dimensional phase space, a bounded hypercubic volume $[-1,+1]^{N_A} \in {\mathbb R}^{N_A}$, and monitoring the system's temporal evolution within that hypercube becomes immediately problematic. Analysis methods based on RPs and RQAs, an approach currently popular and productive in many fields, get us only so far toward our ultimate aim of being able to understand what the system is doing and where it is going (as documented in \cite{alexandrov_2021}) -- and, unfortunately, they do not get us far enough. While it is tempting to invest time and effort in developing better RP/RQA methods for analysis of the PRZI market-system's phase-space trajectories in its subspace of ${\mathbb R}^{N_A}$, the results we present in the next section cast doubt on whether that would actually be a useful thing to do. There,
we discuss the consequences of taking a second small step in the direction of greater realism: one in which the space of possible strategies is still infinite, but is also {\em unbounded}. Once we get there, RP/RQA analysis totally runs out of steam.

\section{Coevolution in an Unbounded Infinite Space: STGP}
\label{sec:STGP}

While the work discussed in the previous section is illuminating, our PRZI-market model can be criticised for its lack of realism in the sense that each adaptive PRZI trader is constrained to play a zero-intelligence strategy that is either ZIC, GVWY, SHVR, or some intermediate hybrid mix: traders in the coevolving PRZI market are never even going to play a more sophisticated minimal-intelligence strategy like AA, GDX, or ZIP. But our work is motivated by the observation that in real-world coevolving markets, the trading entities are not constrained to select between a fixed number of existing pure strategies, and nor are they constrained to choose a point in some continuous subspace that includes specific pure strategies as special cases. In real markets, any entity at any time is free to invent its own strategy or to alter/extend an existing one. Our work with PRZI has revealed some of the issues of visualising and analysing such systems, but the bounded nature of its ${\mathbb R}^{N_A}$ subspace means that it can never show the kind of coevolutionary dynamics of the class of system that we seek to ultimately address in our work. Thus, we need a model in which the space of strategies is not only infinite but also {\em unbounded}. In this section, we briefly describe early results from ongoing work in which each entity does have the freedom to adapt by innovating, by creating wholly new strategies, and in which the space of possible strategies is unbounded and hence infinite. 

Genetic Programming (GP: see e.g.\ 
\cite{koza_1993,poli_etal_2008})
is a form of evolutionary computing in which a genetic algorithm operates on `genomes' that are encodings of programs in a list-based functional language such as 
{\sc Lisp} \cite{touretzky_2013_lisp}  or {\em Clojure} \cite{miller_etal_2018_clojure}. 
Starting with an initial population of programs ${\cal P}_1$, each of the $N_p$ individuals $i$ in ${\cal P}_1$ is evaluated via a {\em fitness function}\/ which assigns a scalar {\em fitness} value $f_i(t)$ to that individual. When all $N_p$ individuals in ${\cal P}_1$ have been evaluated and assigned a $f_i(t)$ value, a new population of $N_p$ individuals is created by a process of {\em breeding}\/ where pairs of individuals in ${\cal P}_1$ are selected with a probability proportionate to their fitness (so fitter individuals are more likely to be selected for breeding) and one or more {\em children} are created that have genomes which inherit from the pair of {\em parents} in ways inspired by real-world sexual reproduction with mutation. In this way, the population of new children ${\cal P}_2$ becomes the next {\em generation} of the system; the old population ${\cal P}_1$ is typically discarded, each individual in ${\cal P}_2$ then has its fitness evaluated, and the next generation ${\cal P}_3$ is then bred from ${\cal P}_2$'s fitter members: if this process is repeated for sufficiently many generations, and if hyperparameters such as the mutation rate are set correctly, then useful novel programs can be created by the `Blind Watchmaker' \cite{dawkins_1986} of Darwinian evolution.  

\begin{figure}
 \begin{center}
\includegraphics[width=0.7\textwidth]{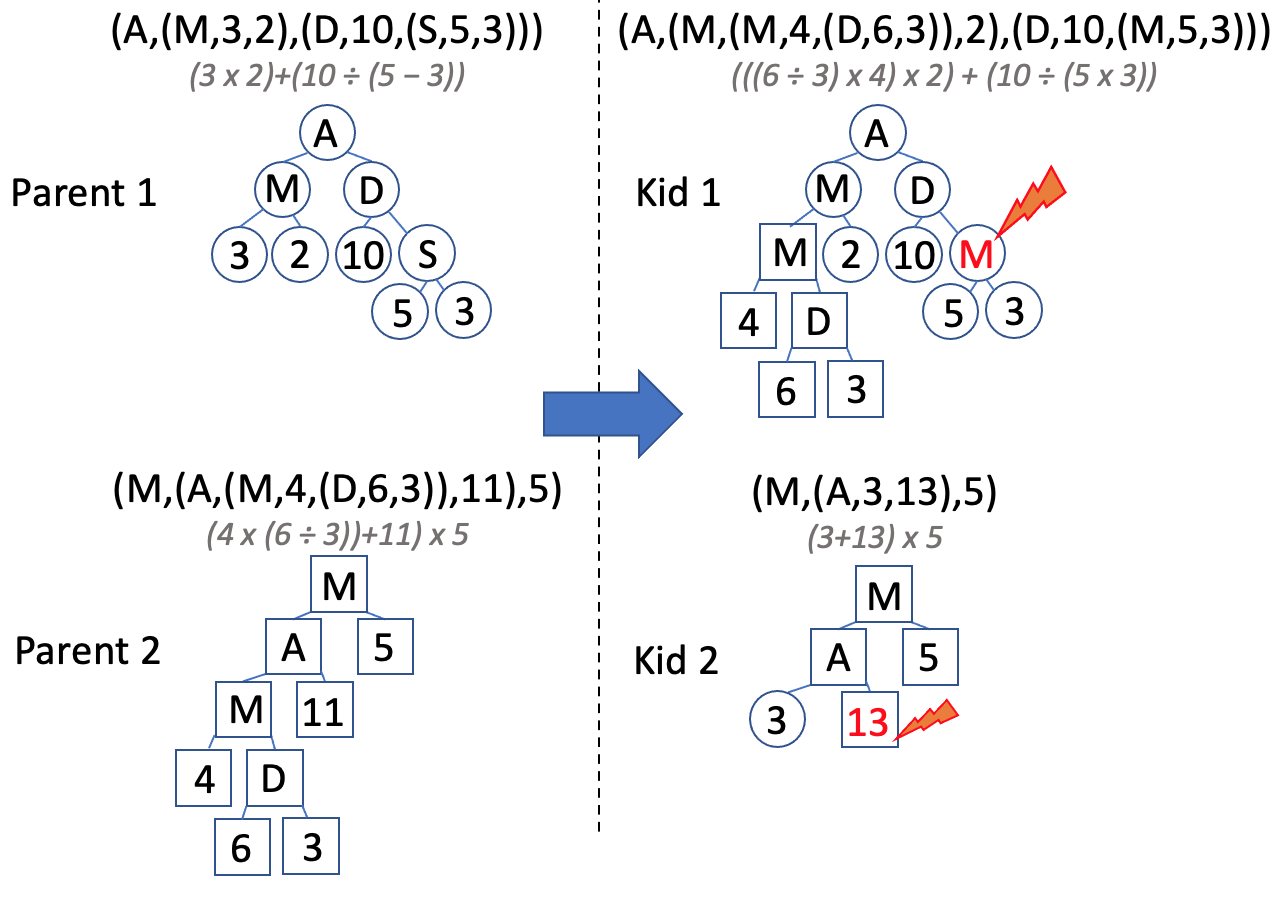}
\end{center}
 \caption{In Genetic Programming, expressions or programs in a list-based functional language are evolved via sexual reproduction and mutation. Here we see genomes for simple mathematical expressions, evaluable on a four function calculator (A is addition; D is division; M is multiplication; S is subtraction). For each genome we show the list representation of the expression, under that the infix mathematical expression in italic font, and below that the tree diagram for the list. On the left we see two parents selected for breeding; on the right we see their two kids, child genomes formed by {\em crossover} (in which a randomly-chosen subtree on one parent is swapped for a randomly-chosen subtree on the other, and {\em vice versa}), and {\em mutation} (indicated by the lightning-flash icons, where the value at a randomly-chosen node in the tree is switched to some randomly chosen other value that is valid at that node). As is shown here, the child genomes can be either longer or shorter than those of their parents. And, if longer genomes encode more sophisticated algorithms that confer greater fitness, then in principle the sophistication of the programs encoded on the genomes can increase indefinitely.}
\label{fig:trees}
\end{figure}

To illustrate this, consider a simple functional language that allows for expressions computable by a four-function pocket calculator, where multiplication has the symbol $M$, division has $D$, subtraction $S$, and addition $A$. The expression $(3 \times 2) + (10 \div (5 - 3))$ (which evaluates to 11) could be written in a list-based style as $(A, (M, 3, 2), (D, 10, (S, 5,3)))$, and can be visualised as a tree structure, as illustrated in Figure~\ref{fig:trees}, which also illustrates the breeding process. Although we have shown only simple mathematical expressions here, when GP is used with Turing-complete languages such as {\sc Lisp} or Clojure, complete executable programs of arbitrary complexity and sophistication can in principle be evolved. 
 
In our work we are using a variant of GP known as strongly-typed genetic programming (STGP), where data-type constraints are enforced between connecting nodes of a program trees \cite{montana_1995}. 
For example, an {\sc and} node that takes two boolean inputs can be guaranteed that it will only connect to two booleans.
Now each entity in our model market, rather than using the Adaptive Climber algorithm to optimize a single numeric $s_i$ strategy value, instead uses a STGP process to create new programs that implement trading strategies: we start with a population seeded with minimally simple programs, and then we unleash them, allowing the coevolutionary process to proceed, during which each entity is at liberty to create programs of growing complexity and sophistication, if in doing so they generate greater profits.   

Full details of our STGP work are given in \cite{figuero_2021}, to which the reader is referred for further detail; here we present only the briefest of results, from a single successful experiment, to motivate discussion of the problems of visualization and analysis that arise when working in this unbounded infinite space of possible programmatic trading strategies.  

As an initial exploration into the dynamics of the STGP traders coevolving in BSE, a simulation was run over 40 generations for 10000 units of time. 100 ZIC sellers were run against 50 ZIC and 50 STGP buyers; both buyers and sellers were regularly replenished with fresh ``customer orders'' (i.e., an instruction to buy or to sell, and an associated private limit price for that transaction) to execute. The STGP traders were each initialised with a price-improvement expression of $(S(S(P^*_{{\text same}}, 1), \lambda_{i,c})$, where $S$ is the subtraction operator, $P^*_{{\text same}}$ is the best price on the same side of the LOB as the trader, and $\lambda_{i,c}$ is the limit price for this customer order $c$ to be executed by trader $i$. This expression represents the zero-intelligence SHVR trader, expressed in STGP tree form.

\begin{figure}[h]
\begin{center}
\includegraphics[width=0.55\textwidth]{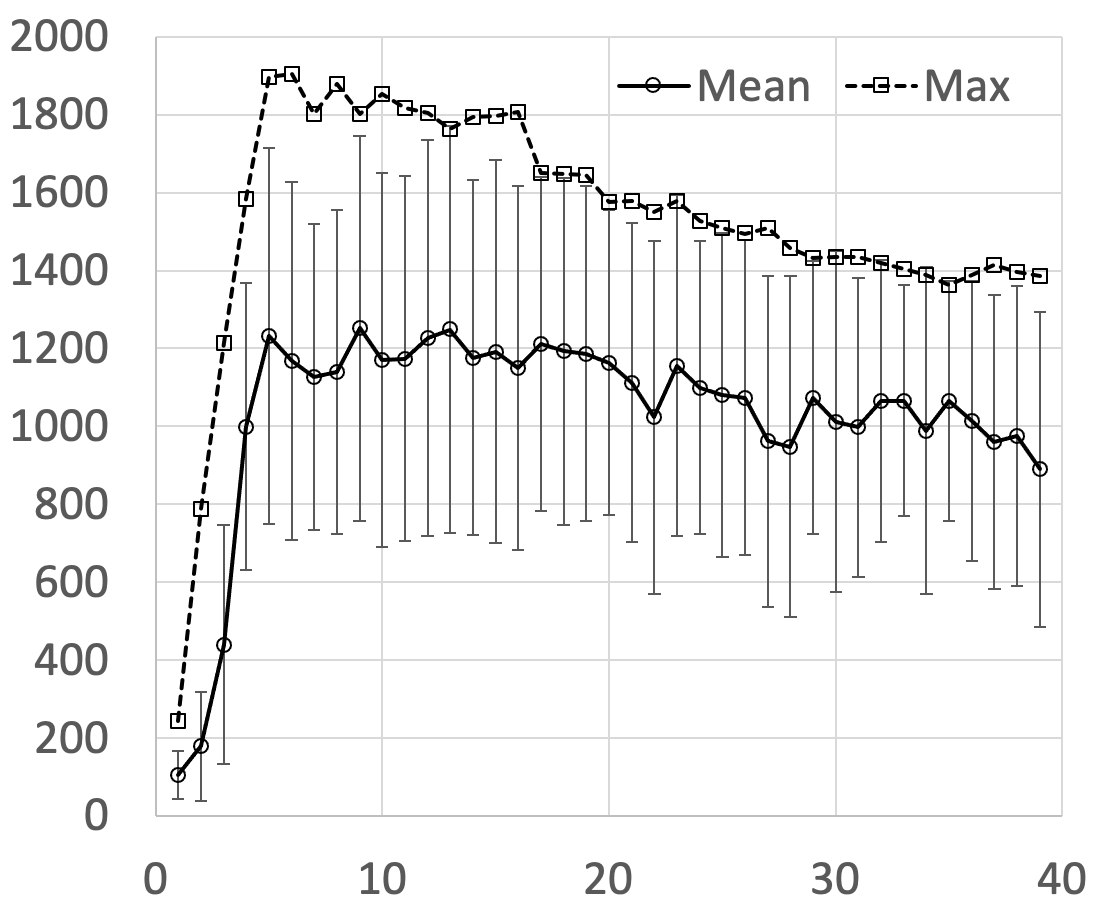}
\end{center}
\caption{Results from an illustrative successful STGP experiment: horizontal axis is generation-number; vertical axis is profit. At each generation the maximum profit achieved by a STGP trader is plotted, along with the mean profitability of all 50 STGP traders; error-bars show $\pm 1$ standard deviation around the mean.}
\label{fig:STGPresults}
\end{figure}

Summary results, a plot of profit-values in each generation, are shown in Figure~\ref{fig:STGPresults}. As can be seen, the profitability data are biphasic: there is an initial brief phase of rapid growth in profitability; followed by a prolonged phase where profitability steadily declines. The initial rise in profitability is as would be expected, and hoped for: the STGP coevolution is discovering ever more profitable trading strategies over successive early generations. The second phase, where profits are steadily eroded, is perhaps less expected and less desired, but can readily be explained by the competitive coevolutionary process progressively eating away at profits: if one SHVR-like trader is profitable by shaving $2\cent$ off of the best price on each revision, then it can be beaten to the deal by another SHVR-like trader who instead shaves $3\cent$; but {\em that}\/ trader could in turn be beaten by a SHVR-style trader who instead shaves $4\cent$ off the best price, and so on:  price-competition among the coevolving traders awards higher fitness to those individuals that get more deals by shaving greater amounts off of the current best price on the LOB, but in doing so the most successful cut their margins ever smaller, eventually hitting a zero margin at which point they are playing not SHVR but GVWY.

Table~\ref{tbl:genomes} shows the genome of the elite (most profitable) trader in a selection of generations from the experiment illustrated in Figure~\ref{fig:STGPresults}. There are two things to note in the genomes shown here. First, STGP (and vanilla GP too) frequently suffers from {\em bloat}, creating viable expressions or programs that get the job done, but which are expressed in very verbose form: for example, the elite individual at generation 30 has a genome that translates to $P^*_{\text best} -1-7-7-1-1-7-1-1-7-1$, which any competent programmer would immediate rewrite as $P^*_{\text best} -  34$ (i.e., as a shortened genome of $(S, P^*_{\text best}, 34))$. Second, because the functional languages used in (ST)GP are richly expressive (that is, the same algorithm or expression can be written in many different ways), the use of methods based on recurrence plots (RPs) becomes deeply problematic: the recurrence of any one particular strategy that had occurred earlier in the evolutionary process may be difficult to automatically detect. For instance, if the elite genome is $(S, P^*_{\text best}, 34)$ at generation 30, and is $(S, (S, P^*_{\text best}, 14), 20)$ at generation 60, and is $(S, (S, P^*_{\text best}, 44), -10)$ at generation 60, then we humans can see by inspection that the same strategy is recurring every 30 generations, but an automated analysis technique would need to go beyond the lexical/syntactic dissimilarity in these expressions and instead reason about the underlying {\em semantics} of the functional programming language. For the simple mathematical expressions being discussed here, it is reasonable to operationally reduce them each to some agreed canonical form, but for only slightly more sophisticated (and stateful) algorithms such as AA, GDX, or ZIP, a many-to-one mapping, a reduction of all possible implementations, all possible expressions, of that algorithm down to a single canonical form is unlikely to ever be achievable. And so, RP-based methods cease to have any applicability here too. 

Once
again, we take a small step in the direction of increased realism in our coevolutionary models, and the visualization/analytics tool-box is empty. 

\begin{table}
\begin{center}
\begin{tabular}{ c c}
 \hline
Gen & Expression Tree\\
 \hline
 1 &  (S,(S,$P^*_{\text best}$,1), $\lambda_{i,c}$)\\
 2 &  (S,(S,$P^*_{\text best}$,1),1)\\
 3 &  (S,(S,$P^*_{\text best}$,1),1)\\
 4 &  (S,(S,(S,$P^*_{\text best}$,1),1),1)\\
 \vdots & \vdots \\
 26 & (S,(S,(S,(S,(S,(S,(S,(S,(S,$P^*_{\text best}$,1),7),1),1),7),1),7),7),1)\\
 27 & (S,(S,(S,(S,(S,(S,(S,(S,(S,$P^*_{\text best}$,1),7),1),7),7),1),1),7),1)\\
 28 & (S,(S,(S,(S,(S,(S,(S,(S,(S,$P^*_{\text best}$,1),7),1),1),7),1),7),7),1)\\
 29 & (S,(S,(S,(S,(S,(S,(S,(S,(S,(S,$P^*_{\text best}$,1),7),7),1),1),7),1),1),7),1)\\
 30 & (S,(S,(S,(S,(S,(S,(S,(S,(S,(S,$P^*_{\text best}$,1),7),7),1),1),7),1),1),7),1)\\
 \hline
\end{tabular}
\end{center}
\caption{Selected STGP genomes for the best individual in the population at various generations in the experiment illustrated in Figure~\ref{fig:STGPresults}: see text for discussion.}
\label{tbl:genomes}
\end{table}

\section{Discussion and Conclusion}
\label{sec:conclusion}

The experiments and results that we have described here have demonstrated that, when we move our ACE-style market models ever so slightly in the direction of being closer to real-world markets, we find that the toolbox for visualisation and analysis of the resultant system dynamics starts to look very empty. While it is relatively easy to make the changes necessary to extend existing models to make them more realistic, it is relatively hard to work out what the extended systems are actually doing, and hence we need new tools to help us do that. Our current work is concentrated on exploring the use of {\sc Ciao} Plots \cite{cliff_miller_1995,cliff_miller_2006} 
in characterising the coevolutionary dynamics of our STGP 
system, although as \cite{cartlidge_bullock_2004} discuss, this is a visualisation technique that is not without its complexities. 

While many research papers in science and engineering are written to describe the solution to some problem, this is not one of those papers. Instead, this is a paper that describes a problem in need of a solution. Or, more specifically, a problem that we expect to be tackled from multiple perspectives, one that eventually yields to multiple complementary solutions. In future work, we intend to develop novel visualisation and analysis techniques for coevolutionary market systems with unboundedly infinite continuous strategy spaces, which we will report on in due course; but in writing this paper we hope to encourage other researchers to work on this challenging problem too. To facilitate that, we have made our Python source-code freely available as open-source releases on GitHub, which is where in future we will also release our own visualisation and analysis methods as we develop them.\footnote{The Python code in the main BSE GitHub repository \cite{cliff_2012_bse} has been extended by addition of a minimally simple adaptive PRZI trader, a $k$-point stochastic hill climber, referred to as PRZI-SHC-$k$ (pronounced {\em prezzy-shuck}), for which the $k=2$ case is a close relative of the AC algorithm described in Section~\ref{sec:PRZI} and which can readily be used for studies of coevolutionary dynamics. The source-code for our STGP work is available separately at {\tt 
https://github.com/charliefiguero/stgp-trader/}.}

%
%

\bibliographystyle{splncs_srt}
\bibliography{../dc_bibliography}

\end{document}